# Cognitive Process of Comprehension in Requirement Analysis in IT Applications

Abhishek Kotnala[1] , R Selvarani [2]

[1] 4th Year B.E Student, Dept. of CSE
Dayananda Sagar College of Engineering, Bangalore-78
Email: kotnala.abhishek@gmail.com
[2] Professor, Dept. of CSE
Head – Software Engineering Research Interest Group, RIIC
Dayananda Sagar College of Engineering, Bangalore-78
selvss@yahoo.com

## ABSTRACT

Requirement Analysis is an important phase in software development which deals with understanding the customers requirements. It includes the collection of information from the customer, which is regarding the customers requirements and what he expects from the software which is to be developed. By doing so, you can have a better understanding of what the customer actually needs and hence can deliver the output as per the customers requirements. Studies are being carried out to bring about improvements in the process of requirement analysis so that errors in software development could be minimized and hence improved and reliable products could be delivered. The human brain comprises of many processes such as thinking, planning, memorizing and analyzing. The brain tends to remember almost everything what it has already experienced in the past. This pattern which gets stored in the memory can be retrieved and used in numerous applications. This paper intends to describe how existing knowledge which has been previously gained and was stored in the brain plays an important role in the requirement analysis process and helps to understand the customers needs effectively and minimize errors and deliver an improved and refined software product.

# I. INTRODUCTION

The human brain has always been a mysterious and unexplored entity with respect to its numerous functions and processes. The brain tends to capture and store whatever it comes across and remembers it for a long time [1] [5]. This previous knowledge which is gained from experiences in the past can prove to be very useful for a variety of applications. In requirement analysis phase of software development process, we can make use of our past knowledge (history) and thus greatly improve the final product. It is well known that the process of requirement analysis is extremely vital in the software development process and hence proper attention should be given to it [3] [8].

This paper discusses how one of the cognitive processes, namely, comprehension plays an important role in the requirement analysis phase. For the organization of the paper, section II discusses about the background of requirement analysis and cognitive process of comprehension. The role of comprehension in requirement analysis is discussed in section III and the final section IV gives a brief conclusion.

# II. OVERVIEW: REQUIREMENT ANALYSIS AND COGNITIVE PROCESS OF COMPREHENSION

## A. Requirement analysis

It is the phase in software development which basically deals with understanding the needs of the customer [3] [8]. It helps in understanding "what" the application, which is to be designed is expected to do. This phase includes collection of data from the customer regarding his requirements and then modeling and analyzing these requirements in order to form a basis for software design [8]. This phase is extremely vital in the process of software development because, if adequate attention is not given to it then there is a possibility that the developed software may not meet the customers' requirement or may face complete failure.

The process of requirement analysis includes the following phases [3]

- Domain Understanding
- Requirements Collections
- Requirements Classification
- Setting Priorities
- Resolving Conflicts
- Requirements Validation

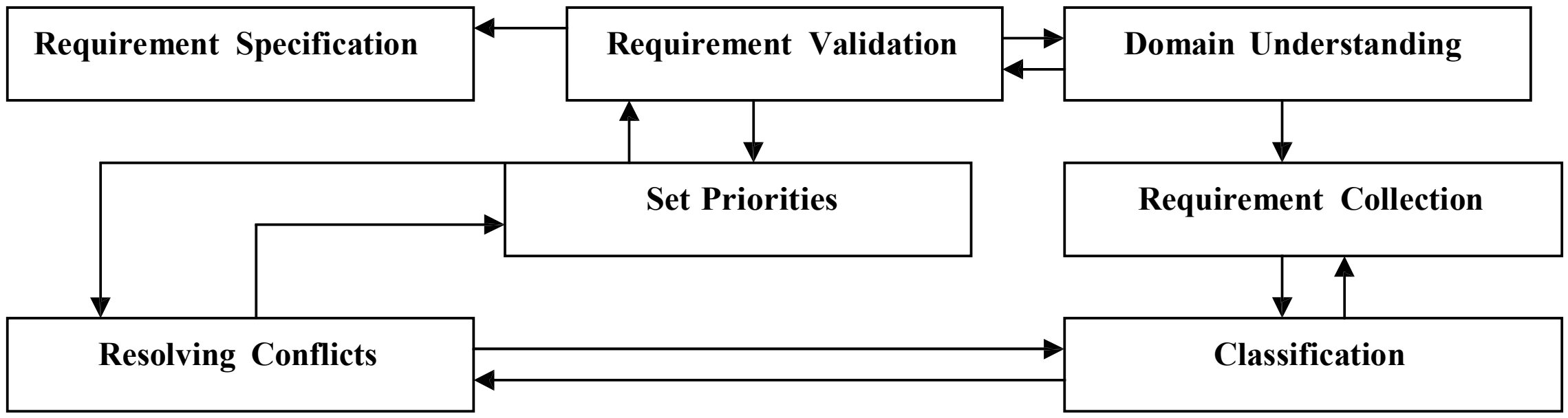

Figure. 1 Process of requirement analysis (reference Dr. Jerry Gao)

B. Cognitive process of comprehension

Cognition can be broadly described as an interaction between knowledge driven processes and sensory processes and between controlled processes and automatic processes. The process of cognition includes various states and processes of the mind such as thinking, planning and reasoning [1] [2] [6].

Comprehension is defined as the capability of understanding something. It is a process of the brain in which internal relations are formed between various objects and attributes. At later stages, information is extracted from this internal representation. The brain forms a relational OAR model based upon these relations between the objects and attributes [1].

In the cognitive process of comprehension the brain tries to retrieve a known fact which is stored in memory.

The steps involved in the process of comprehension are [1]:

1. identify the given object
2. find relations between related objects which are already stored
3. if the findings are adequate- build a OAR model
4. else- search external resources
5. if sufficient relations are found, then store the results in the memory

## III. ROLE OF COGNITIVE PROCESS OF COMPREHENSION IN SOFTWARE REQUIREMENT ANALYSIS

In the case of requirement analysis the comprehension process begins by identifying the requirement of the customer which acts as the input object. When the object is identified, the brain searches for the relations between the input object and the objects already existing in the brain. For requirement analysis, this happens by relating the customer's requirement with any information which the system analyst had come across in the past and it was stored in his memory. The brain will search for any similar objects or attributes which could be related to the current input object, that is, the present requirement of the customer. This is done by identifying similar actions which had been performed in the past and can provide a better understanding of the customer's requirement. If the relations and findings found in this manner are adequate then the brain proceeds to construct an internal representation of all the related data which is called the OAR model for the given input requirement of the customer. This generally would contain relations between the present input and the past experience.

For example, suppose the customer puts forward his requirement in front of the system analyst. Now, while analyzing the data collected from the customer the analyst will first think about many of his past experiences which would help him to get a better understanding of the complexity of the problem. Here, the brain would start relating the present problem with all the previously gained data in the past. By doing so, the analyst will be careful about all the actions and decisions taken by him in the past when dealing with a similar problem. If he had made mistakes in the past, he would make necessary attempts not to repeat them again or if there had been some important steps, they could be applied again to get better results. But if the findings are not sufficient enough, the brain will search for any further information from external resources like if the system analyst is unable to comprehend the customer's requirement properly or he may not be able to think of any past experience which could be of help then in such a case the analyst would prefer to go for external resources such as referring to relevant books, collecting information from the internet or maybe talking to people about their experiences and then finally be able to get a better perspective of the problem. In this case the brain constructs a partial OAR model [1].

## IV. CONCLUSION

This paper has presented how the cognitive process of comprehension, which is, the ability of the brain to understand and relate things, plays an important role in the requirement analysis process of software development. This process of comprehension thus leads to better understanding of the customers' requirement and hence leads to improved results. The finding of this paper has also indicated that the next generation computer memory systems may be built according to the traditional container metaphor which is more powerful ,flexible efficient, and is capable of generating a mathematically unlimited memory capacity by using limited number of neurons in the brain of hardware cells in the next generation computers.